# Temperature-Aware Scheduling of LLM Inference in Large-Scale Geo-Distributed Edge Data Centers with Distributed Optimization

Arash Khalatbarisoltani[1], Amin Mahmoudi[2], Jie Han[3], Muhammad Saeed[4], Wenxue Liu[1], Jinwen Li[1], Solmaz Kahourzade[5], Amirmehdi Yazdani[6], Xiaosong Hu[1]

**Abstract**—The environmental impact of Large Language Models (LLMs) on data centers hosting these models is becoming a significant concern. While many efforts have focused on reducing the substantial training overhead of LLMs, carbon and water consumption during the inference phase can often surpass the costs associated with their training. The cooling systems of data centers are crucial in this context, but they are frequently modeled with a location-independent efficiency term. However, their energy efficiency is highly influenced by ambient temperature, which can vary significantly across different geographical locations. Leveraging this temperature diversity can help reduce total cooling energy costs and improve the performance of edge data centers. To address these critical sustainability issues related to LLMs, this study proposes a temperature-aware approach that co-optimizes LLM energy costs, carbon emissions, time-to-first token, and water consumption. The approach employs a distributed optimization algorithm based on an alternating direction method of multipliers, aimed at enhancing the sustainability of LLM hosting across geo-distributed edge data centers in Australia. Our method demonstrates reductions in cooling energy consumption and improves overall cost efficiency for geo-distributed cloud environments.

## 1. Introduction

The growth of Large Language Models (LLMs) is increasingly dominating cloud computing workloads. Over the past few years, LLM usage has nearly doubled, leading to a significant rise in energy consumption among cloud servers, which raises major concerns. While LLM training is often viewed as the primary issue, it is surprising to note that the inference phase of LLMs actually dominates costs, consuming an estimated 25 times more computational resources annually than the training phase [1]. The carbon footprint associated with LLM inference is another critical consideration, with estimates suggesting that it can be 1,400 times greater than that of training for large-scale data centers each year [2]. Furthermore, the environmental impact of LLMs is exacerbated by significant water consumption. Data centers consume vast amounts of water—illustratively, a medium-sized data center can use as much water as two 18-hole golf courses [3]. LLM inference is contributing to the increasing water consumption in these edge data centers, a trend that must be addressed. Workload scheduling within data centers has been extensively studied, focusing mainly on energy efficiency and performance (latency) as key metrics. For instance, [4] proposes a game-theoretic approach that considers data transfer costs to minimize energy consumption, while [5] introduces a hybrid genetic algorithm aimed at reducing workload latency. As climate change becomes an urgent issue, various recent solutions have emerged to mitigate the environmental impact of LLMs. For example, [6] suggests a method that improves both carbon emissions and throughput. Some recent efforts also factor in water consumption alongside carbon emissions, such as the hybrid scheduling heuristic proposed in [7], which targets complex geo-distributed networks of data centers. To address the unique challenges of managing LLM inference, innovative workload management strategies are necessary. The adaptive method presented in [8] aims to balance training loads across multiple heterogeneous data centers; however, this approach is not suitable for inference workloads that require low-latency processing close to end-users. Meanwhile, [9] offers a mixed-integer linear programming method that effectively balances requests across heterogeneous hardware. To enhance throughput in terms of tokens per second, [10] proposes a method that splits each LLM request into two phases. With respect to the energy footprint of data centers, cooling systems—which account for 30% to 50% of total energy usage [11]—are often modeled using a location-independent efficiency value. However, [12] highlights that ambient temperature significantly influences cooling energy consumption, particularly in data centers utilizing outside air cooling systems. As demonstrated by [12], when the ambient temperature decreases from 35°C to -3.9°C, the partial Power Usage Effectiveness (PUE) of data centers shifts from 1.30 to 1.05, indicating that cooling energy consumption varies by location and time. This underscores the importance of implementing temperature-aware workload management for LLMs, leveraging temperature variations alongside other factors to reduce overall cooling energy overhead in geo-distributed edge data centers. This paper proposes a temperature-aware optimization approach for scheduling LLM inference requests in geo-distributed environments, aimed at decreasing energy costs, reducing carbon emissions, enhancing time-to-first token (TTFT), and minimizing water consumption. The proposed distributed approach employs an optimization method based on the alternating direction method of multipliers (ADMM). The key contributions of this study are as follows: A temperature-aware distributed optimization approach for LLM inference workloads tailored to geo-distributed edge data centers in Australia. A formulation of the LLM inference scheduling problem that co-optimizes energy costs, carbon emissions, TTFT, and water consumption for each LLM inference request. A comprehensive model of carbon, water consumption, and energy costs for each data center, accounting for varying temperatures across heterogeneous LLM-hosting data centers. A comparison of the proposed temperature-aware scheduling approach with several existing methods, demonstrating its capacity to achieve superior solutions. The remaining sections of the paper are organized as follows: Section 2 presents the models utilized; Section 3 outlines the suggested approach; Section 4 presents the results; and Section 5 provides concluding remarks.

[1]Chongqing University, Chongqing, China
[2]Flinders University, Adelaide, Australia
[3]Loughborough University, Loughborough, England
[4]Shanghai Jiao Tong University, Shanghai, China
[5]University of South Australia, Adelaide, Australia
[6]Murdoch University, Perth, Australia

## 2. Datacenter model

Each LLM edge data center comprises of $G_s$ nodes at each data center site $s$. Each computing node $g \in G_s$ involves multiple GPUs. Energy consumption is estimated on a per-node basis with each computing node $g \in G_s$ have working states $p \in [ON, IDLE, OFF]$ with each working state is a proportion $PR_p$ of the thermal design power $TDP_g$ of the computing node [13]. Therefore, the information technology (IT) energy consumption of each computing node $E_{g,t}^{IT}$ can be calculated by

$$E_{g,t}^{IT} = t \cdot PR_p \cdot TDP_g, \tag{1}$$

The total IT device energy consumption $E_{l,t}^{IT}$ across all computing nodes $G_l$ at site $s$ can be calculated by
$$E_{s,t}^{IT} = \sum_g^{G_s} E_{g,t}^{IT}, \tag{2}$$

Regarding the mechanical cooling system energy consumption $E_{s,t}^{M}$, the largest portion is associated with the energy consumption of computer room air conditioning (CRAC) $E_{s,t}^{C}$ and is determined by
$$E_{s,t}^{C} = \frac{E_{s,t}^{IT}}{CoP_s}, \tag{3}$$

where $CoP_s$ denotes the cooling system efficiency. The chillers and the other hardware consume energy usage that is approximately with the same amount of CRAC units $E_{s,t}^{C}$ [14]. Then, the whole cooling energy consumption $E_{s,t}^{M}$ can be calculated by
$$E_{s,t}^{M} = 3 \cdot E_{s,t}^{C}, \tag{4}$$

The energy usage of the power conditioning unit $E_{s,t}^{S}$ is equivalent to about 13% [15] of the whole IT device energy consumption $E_{s,t}^{IT}$. Finally, the total energy consumption $E_{s,t}^{T}$ of each edge datacenter s can be estimated by
$$E_{s,t}^{T} = E_{s,t}^{IT} + E_{s,t}^{M} + E_{s,t}^{S}, \tag{5}$$

The energy consumption cost is dependent on how the utility company charges for consumption on the power grid. The cost due to different pricing around peak hours and the generation approach (e.g., coal, hydroelectric, solar) are added by time-of-use price $TOU_{s,t}$. Ultimately, the edge datacenter energy cost $P_t^T$ at site $s$ can be calculated by

$$P_t^T = \sum_s^S (E_{l,t}^T \cdot TOU_{l,t}), \quad (6)$$

Generally, water consumption in an edge datacenter occurs because of the cooling unit on-site and from the power generation off-site. The water consumptions for an edge datacenter $s$ involve the evaporative water consumption $W_s^E$, blowdown water consumption $W_s^B$, and grid water consumption $W_s^G$. The evaporative part of water consumption $W_s^E$ can be calculated by

$$W_{s,t}^E = \frac{H_{s,t}^{IT}}{H_w} \quad (7)$$

where $H_w$ represents the heat capacity of water. Each mechanical cooling unit of data centers involves a coolant (typically potable water), but the water of this unit must be handled at a wastewater facility unit [16]. The blowdown water $W_{s,t}^B$ can be formulated by:

$$W_{s,t}^B = \frac{W_{s,t}^E}{1-D}, \quad (8)$$

where $D$ is the ratio of the water due to the buildup of solid chemicals. In addition to that, each power source has a water intensity $WI_s$. For instance, the water intensity $WI_s$ of wind and hydropower are 0.2 and 67 L/kWh, respectively [17]. Therefore, the total water intensity of the grid for edge datacenter $s$ can be calculated by

$$W_{s,t}^G = E_{s,t}^T \cdot WI_{s,t}, \quad (9)$$

Therefore, the whole water consumption per LLM interference epoch $W_t^T$ can be calculated by

$$W_t^T = \sum_s^S (W_{l,t}^E + W_{l,t}^B + W_{l,t}^G), \quad (10)$$

The carbon emissions model takes into consideration both the costs associated with electricity and water. The grid power carbon emission $C_{s,t}^G$ is formulated by

$$C_{s,t}^G = CI_{s,t} \cdot E_{s,t}^T, \quad (11)$$

where $CI_{s,t}$ represents the carbon intensity of the grid $CI_{s,t}$ at site $s$. Additional parts of carbon emission come from the potable water generation $EI_{p,t}$ and also the wastewater processing $EI_{w,t}$. Then, the carbon emissions directly associated with water consumption can be estimated by

$$C_{s,t}^W = CI_{s,t} \cdot [(W_{s,t}^B + W_{s,t}^E) \cdot EI_{p,t} + W_{s,t}^G EI_{w,t}], \quad (12)$$

Ultimately, the total carbon emissions over any epoch can be calculated by

$$C_t^T = \sum_s^S (C_{s,t}^G + C_{s,t}^W), \quad (13)$$

The LLM inference requests are modeled by the memory footprint $M_i$ and TTFT. The memory of each LLM inference $M_i$ involves model parameter $M_o$ and the key-value (KV) cache. The KV memory raises per output token $n$ up to the total of all output tokens $N_i$. The total memory $M_i$ can be estimated by

$$M_i = M_o + N_i \cdot M_{o,i}^{KV}, \quad (14)$$

where $M_{o,i}^{KV}$ is the size of the model $o$. The TTFT of each LLM inference must consider the loading overhead $F_o^l$ of the LLM model $o$. The loading orchestration overhead $F_o^l$ is estimated by

$$F_o^l = \frac{M_o}{B_g}, \quad (15)$$

where $M_o$ is the size of the model $o$ and $B_g$ is the bandwidth of the server node $g$.

### 3. Distributed Optimization-based Scheduling

This study focuses on geo-distributed AI edge data centers to host LLM inference workloads. The temperature-aware scheduling approach outlines an execution plan for these workloads, encompassing both workload assignment and scheduling within each data center. The proposed method employs a distributed optimization strategy that utilizes an alternating direction method of multipliers (ADMM). The primary aim is to optimize TTFT, carbon emissions, water consumption, and energy costs.

## 4. Results

This section presents a comparison of the proposed temperature-aware approach with two scheduling methods: mixed-integer linear programming-based (Helix) [9] and queue-based (Splitwise) [10]. Each edge data center has 200 compute nodes available. This study examines 20 data centers located across Australia. Five solutions are considered: Opt-Carbon, Opt-TTFT, Opt-Water, Opt-Cost and balanced solution. [18] is utilized, as shown in Figure 1, which depicts the number of LLM token requests.

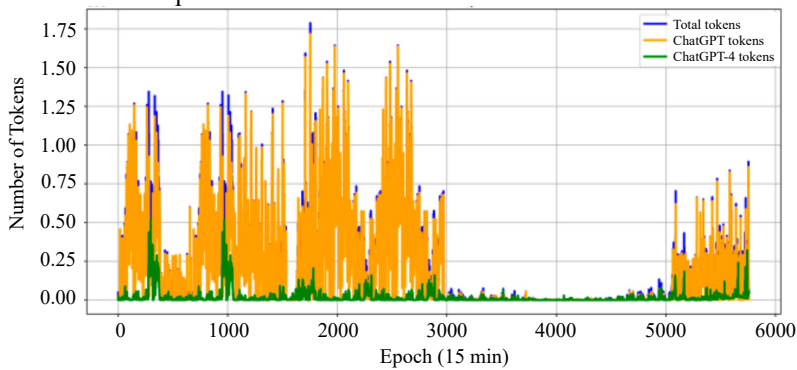

Figure 1. LLM token [18, 19].

Figure 2 illustrates the results of various approaches, normalized against Splitwise's outcomes. Both Splitwise and Opt-Balance achieve comparable TTFT; however, Opt-Balance accomplishes this with reduced carbon emissions, water consumption, and energy costs. Conversely, Helix consistently underperforms compared to Opt-Balance. The results indicate that single-objective optimized solutions surpass both Helix and Splitwise. Opt-Balance provides an effective option, outperforming Helix in all measured metrics. Compared to Splitwise, it not only yields lower carbon emissions, energy costs, and water consumption, but also maintains competitive TTFT performance.

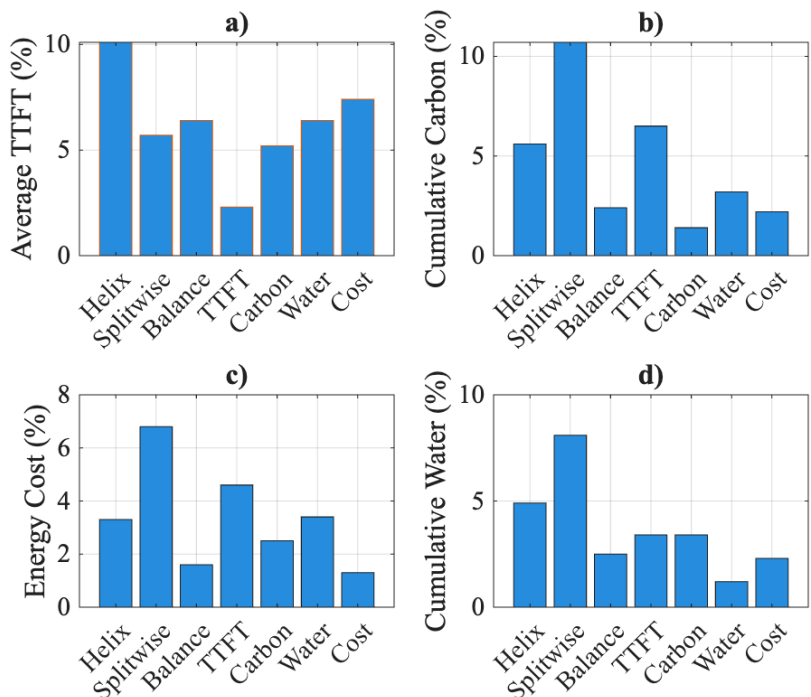

Figure 2. a) TTFT, b) carbon cost, c) power usage cost, and d) water consumption.

## 5. Conclusion

This study introduces a temperature-aware approach to address the challenges of scheduling LLM inference requests across geo-distributed edge data centers. The primary objective is to optimize carbon emissions, water consumption, TTFT, and energy costs while accounting for the impact of temperature. Our experiments revealed that the proposed approach produced promising solutions, surpassing the performance of alternative methodologies.